# Deep learning-based single-shot computational spectrometer using multilayer thin films

**Cheolsun Kim**[1], **Dongju Park**[2], **Jioh Lee**[1], **and Heung-No Lee**[1*]

[1]School of Electrical Engineering and Computer Science, Gwangju Institute of Science and Technology, Gwangju, 61005, South Korea
[2]AI, NAVER Corp., Seongnam 13561, South Korea
*Correspondence: heungno@gist.ac.kr

**ABSTRACT**

Computational spectrometers have mobile application potential, such as on-site detection and self-diagnosis, by offering compact size, fast operation time, high resolution, wide working range, and low-cost production. Although these spectrometers have been extensively studied, demonstrations are confined to a few examples of straightforward spectra. This study demonstrates deep learning (DL)-based single-shot computational spectrometer for narrow and broad spectra using a multilayer thin-film filter array. For measuring light intensities, the device was built by attaching the filter array, fabricated using a wafer-level stencil lithography process, to a complementary metal-oxide-semiconductor image sensor. All the intensities were extracted from a monochrome image captured with a single exposure. A DL architecture comprising a dense layer and a U-Net backbone with residual connections was employed for spectrum reconstruction. The measured intensities were fed into the DL architecture for reconstruction as spectra. We reconstructed 323 continuous spectra with an average root mean squared error of 0.0288 in a 500–850 nm wavelength range with 1-nm spacing. Our computational spectrometer achieved a compact size, fast measuring time, high resolution, and wide working range.

## Introduction

Spectrometers are effective tools used in scientific research and industrial sites for chemical analysis[1], remote sensing[2], etc. Although spectrometers are used in various applications, they are confined to static environments, such as laboratories and factories, due to their bulky size, long operation time, and high cost. Due to these restrictions and practical application requirements, optical filter-based spectroscopy is emerging as a promising technique. A configuration of a filter-based spectrometer could be realized by attaching a filter array to a complementary metal-oxide-semiconductor (CMOS) image sensor. A low-cost and compact design can be realized, unlike grating-based spectrometers that requires diffractive optics and motorized components. However, numerous narrow-shaped bandpass filters are required to cover a wide wavelength range with a high resolution. Fabricating such delicate filters is challenging; it is complicated to integrate them into small CMOS-sensor areas in an array form.

Instead of elaborating optical filters to improve the resolution of spectrometers, computational approaches have been employed to conventional spectrometers to improve resolution[3–5]. Further, various photonic structures[6–20] have been proposed to use advanced computational approaches based on compressive sensing[21]. Unlike conventional spectrometers that selectively measure specific light wavelengths, these photonic structures measure a wide wavelength range with unique transmission functions: thus, they have a good light efficiency. Besides, it is possible to cover a wide wavelength range with a small number. Various types of photonic structures, such as quantum dot filters[9,13], etalon filters[10,14], photonic crystal slabs[15,16], nanowire[17], and multilayer thin films (MTFs)[18,20] have been employed and shown to perform well. However, the reconstruction results are confined to monochromatic lights, light-emitting diodes (LEDs), and laser sources.

Computational spectrometers mostly adopt iterative numerical optimization methods[22–24] based on the constraints that the light sources are sparse signals, or can be sparsely represented by a certain sparsifying basis. However, there is a limitation to representing abundant spectral features with combinations of columns on a fixed sparsifying basis. In addition, these approaches work well for precisely measured signals and handicraft parameters predetermined through prior information, such as spectral sensitivities, sparsifying basis, line shapes, and full width at half maximums (FWHMs) of spectra. Thus, the reconstructing performance of a spectrum could be biased depending on the variation in noise level and predetermined parameters. These limitations make it difficult to use computational spectrometers to recover various waveforms of spectra.

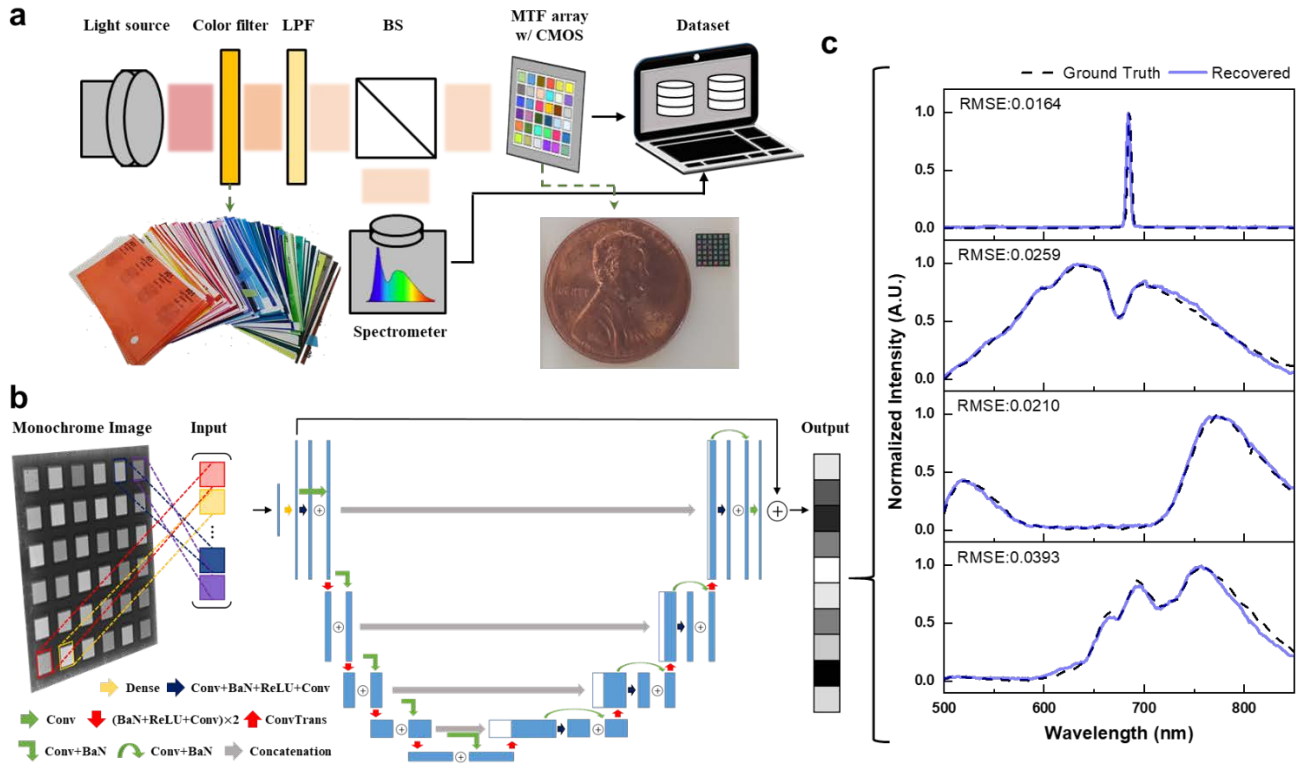

**Figure 1.** DL-based single-shot computational spectrometer. (a) Schematic of the experimental setup. A collimated beam from a light source is split into two beams by a beam splitter (BS) after passing through a color filter and a long pass filter (LPF). The spectrum of a beam is measured using a commercial spectrometer. The other beam is modulated by the MTF filter array and captured as an image by a CMOS camera. (b) DL architecture comprises a dense layer and a U-Net backbone architecture with residual connections. (c) Examples of recovered test spectra using the trained DL architecture. The black dashed lines represent GT spectra measured using a commercial spectrometer. The solid green lines represent reconstructed spectra using the proposed spectrometer.

Recently, deep learning (DL)[25] approaches have been proposed as alternatives to numerical optimization methods for computational spectrometers[26–33]. Kim et al.[27,29] used a convolutional neural network to restore the spectra from sampled intensities. They compared the reconstruction performance of numerical optimization approaches and the convolutional neural network via simulation. Optical experiments to reconstruct light-emitting diodes using a U-Net structure have also been performed[28]. In addition, methods for reconstructed spectra that combine a DL architecture and compressive sensing framework have been proposed[30,31]. Spectral reconstructions using a DL architecture without prior information-demonstrated reconstructions of the combination of narrow bands spectra using a programmable supercontinuum laser have also been demonstrated[32]. Further, hyperspectral imaging via DL-based simultaneous filter design and spectrum reconstruction has been proposed[33].

In this study, we propose a DL-based single-shot computational spectrometer for recovering narrow and broad spectra. As a configuration of the computational spectrometer, we employed an MTF filter array of 36 filters and a CMOS camera. The MTF filter array was fabricated through a wafer-level stencil lithography process, which can be scalable, reproducible, and mass-produced. A computational spectrometer was built by directly attaching the MTF filter array to the CMOS camera. The incident light was modulated by the MTF filter array, where each filter had a unique transmission function. The filtered light was measured by the CMOS camera with a single exposure. From the captured image, we extracted 36 compressively sampled light intensities. The intensities were fed into a DL architecture and reconstructed as a spectrum of 350 elements. We collected 3,223 spectra including narrow and broad spectra for training/validation/testing the DL architecture. After training, we tested the performance using unused data in the training/validation process. The average root mean squared error (RMSE) for the test set was 0.0288. The results show that the proposed DL architecture performed well in simultaneously reconstructing narrow and broad spectra. The proposed spectrometer is compact with a single-shot structure and can be mass-produced. Besides, applying the DL technique can offer a high resolution, wide working range and fast reconstruction. Therefore, the proposed spectrometer can become a new form factor for on-site detection, such as drink inspection, counterfeit document detection, and self-diagnosis.

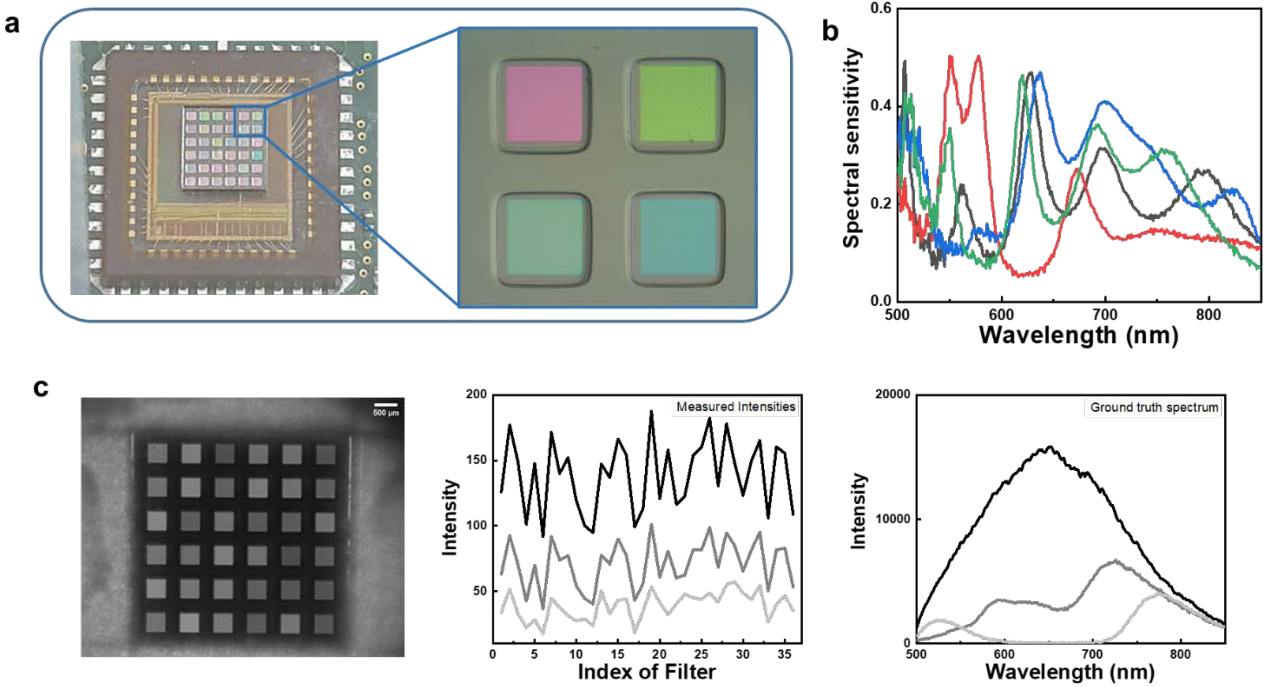

**Figure 2.** Fabricated MTF filter array for single-shot computational spectroscopy. (a) Photograph (left) and optical microscope image (right) of the fabricated MTF filter array. (b) Examples of spectral sensitivities of MTF filters with the CMOS camera. (c) Examples of measured data: a monochrome image of the MTF filter array (right), some extracted intensities from monochrome images (center), and GT spectra of incident lights corresponding to the extracted intensities (right).

## Results

**Principle of DL-based computational spectroscopy.** Figure 1(a) depicts a schematic of our experimental setup. A collimating beam is divided into two beams after passing through a color filter and a long pass filter. A spectrum of one split beam was measured using a commercial spectrometer (Black-Comet, StellarNet), used as the ground truth (GT). The other beam was fed into the MTF filter array and modulated by the transmission functions of the filters. The modulated intensities of the beam were captured with a CMOS camera (EO-1312M, Edmund optics) as a monochrome image with a single exposure. By connecting the spectrometer and CMOS camera to a laptop using universal serial bus cables, we simultaneously collected the monochrome image GT spectrum. The captured monochrome image had a size of 1280 × 1024 pixels, and the GT spectrum comprised a signal of 350 elements measured at the wavelength range, $\lambda$, of 500–850 nm with 1-nm spacing.

As shown in Fig. 1(b), we extracted 36 intensities from the filter array in the monochrome image. These intensities were fed into the DL architecture to reconstruct the spectra. The proposed DL architecture comprises a U-Net backbone architecture with residual connections. We trained the DL architecture using the training set and tested the resolving performance using the test set. Figure 1(c) shows examples of reconstructed test spectra using the trained DL architecture. The RMSE between the GT spectra (dashed black lines) and reconstructed spectra (solid blue lines) was used to evaluate the reconstruction performance $\sqrt{\left\| \mathbf{x}_{recovered} - \mathbf{x}_{GT} \right\|_2^2 / size(\mathbf{x})}$. The reconstructed and GT spectra were consistent, as shown by the RMSE values written in the upper left of each plot (Fig. 1(c)).

**MTF filter array for DL-based computational spectroscopy.** The MTF filter array had a 6 × 6 square grid shape. Each MTF filter had a size of 400 × 400 μm², and the filters were 300 μm apart. The entire size of the filter array was 4.5 × 4.5 mm² (Fig. 2(a)). The filter array was fabricated using wafer-level stencil lithography based on shadow masks, which can be mass-produced, scalable, and reproducible. Figure 2(a) shows a photograph and an optical microscope image of the fabricated MTF filter array. Each MTF filter had its own color due to its unique transmission function, $T$, and the color was uniform across each filter. Filters with unique transmission functions can be generated by stacking multiple layers of thin films with different numbers and thicknesses (see the Methods Section for more details). A single pixel of the CMOS camera had a size of 5.2 × 5.2 μm². Underneath each filter, there were approximately 70 × 70 pixels. An intensity of a pixel underneath the *i*-th filter is defined as the linear relation between the spectral sensitivity, $S_i(\lambda)$, and the target spectrum x($\lambda$), i.e., $I_i(\lambda) = S_i(\lambda)\mathrm{x}(\lambda)$ where $S_i(\lambda)$ can be derived by element-wise multiplication of the transmission function, $T_i(\lambda)$, and the quantum efficiency of the CMOS camera, $QE(\lambda)$. Figure 2(b) shows examples of spectral sensitivities of the fabricated filters with the CMOS camera. Unlike bandpass optical filter-based spectrometers, the MTF filter-based computational spectrometer modulates the spectrum of incident light in a wide wavelength range with broad spectral sensitivities. Therefore, few filters are sufficient to measure the spectral information of the incident light uniquely. Figure 2(c) shows the measured data. From the monochrome image of the MTF filter array illuminated by the incident light (left), 36 measured intensities were extracted by taking the average value of central 40 × 40 pixels of each filter as one measured intensity. For pixels at the filter boundary, there might have been a misalignment during fabrication, and a beam passing through a filter could overlap another beam in experiments. Therefore, we excluded pixels on the filter boundaries. Three examples of measured intensities are plotted in the center of Fig. 2(c), which correspond to the GT spectra on the right of Fig. 2(c).

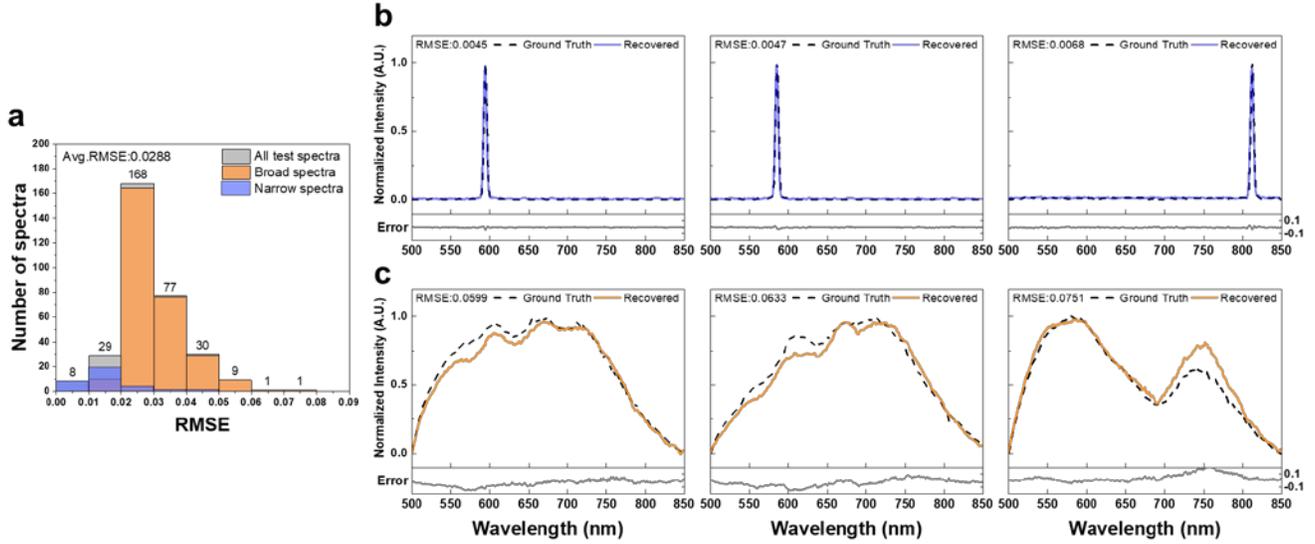

**Figure 3.** Result of spectral reconstructions. (a) Histogram of the RMSE distribution for the test set; the average RMSE is 0.0288. Blue and orange boxes represent the RMSE distribution of narrow and broad spectra, respectively. (b) Three examples of the best spectral reconstructions. (c) Three examples of the worst spectral reconstructions. Dashed black lines represent GT spectra. Solid blue and orange lines represent the reconstructed spectra. Solid light gray lines at the bottom of each graph represent the error.

**DL architecture.** Figure 1(b) depicts the proposed DL architecture schematic comprising a dense layer and a U-Net backbone[34] with residual connections. Before entering the U-Net backbone, 36 measured intensities were extended to a size of 350 by applying a linear transformation using the dense layer. This extension allowed the depth of the U-Net backbone to become deep, which could be helpful in the feature extraction and reconstruction. The U-Net backbone comprises a contracting path and an expansive path. In the first stage of the contracting path, extended intensities go through the main branch that comprises a one-dimensional (1D) convolution (Conv), 1D batch normalization (BaN), rectified linear unit (ReLU), and Conv (navy blue arrow in Fig. 1(b)). As a shortcut branch, the extended intensities go through a Conv. The two branches are added up to be a residual connection. The output of the residual connection becomes the input of the next stage of the contracting path. Like the first stage, the input of the second stage of the contracting path goes through the main and residual branches and becomes a residual connect. In the main branch, we reduce input size by a factor of 2 and double the number of feature maps using two sets of BaN, ReLU, and Conv. We used Conv with stride 2 to reduce the size. In the residual branch, the input goes through Conv with stride 2 and BaN. The output of the contracting path is upsampled by applying a 1D transposed convolution (ConvTrans) and is concatenated with the corresponding feature maps from the contracting path to be the input of the first stage of the expansive path. The input goes through the main and residual branches and is summed. These upsampling, concatenation, and summation were repeated four times. The output of the expanding path goes through a Conv to become a signal with 350 elements. Finally, the output signal and extended intensities were added to become a reconstructed spectrum.

The proposed DL architecture was trained to minimize a mean squared error between the reconstructed and GT spectra. By leveraging the summation between the extended intensities and the output signal of the U-Net backbone, the U-Net backbone learns the *residue* between the extended intensities and ground truth spectra. The learning *residue* is more effective than directly learning target spectra[35]. Residual connections in the U-Net backbone prevent the gradient-vanishing problem, which could stop updating learnable parameters in a DL architecture during the training process[36]. In addition, it is possible to deepen the depth of the DL architecture.

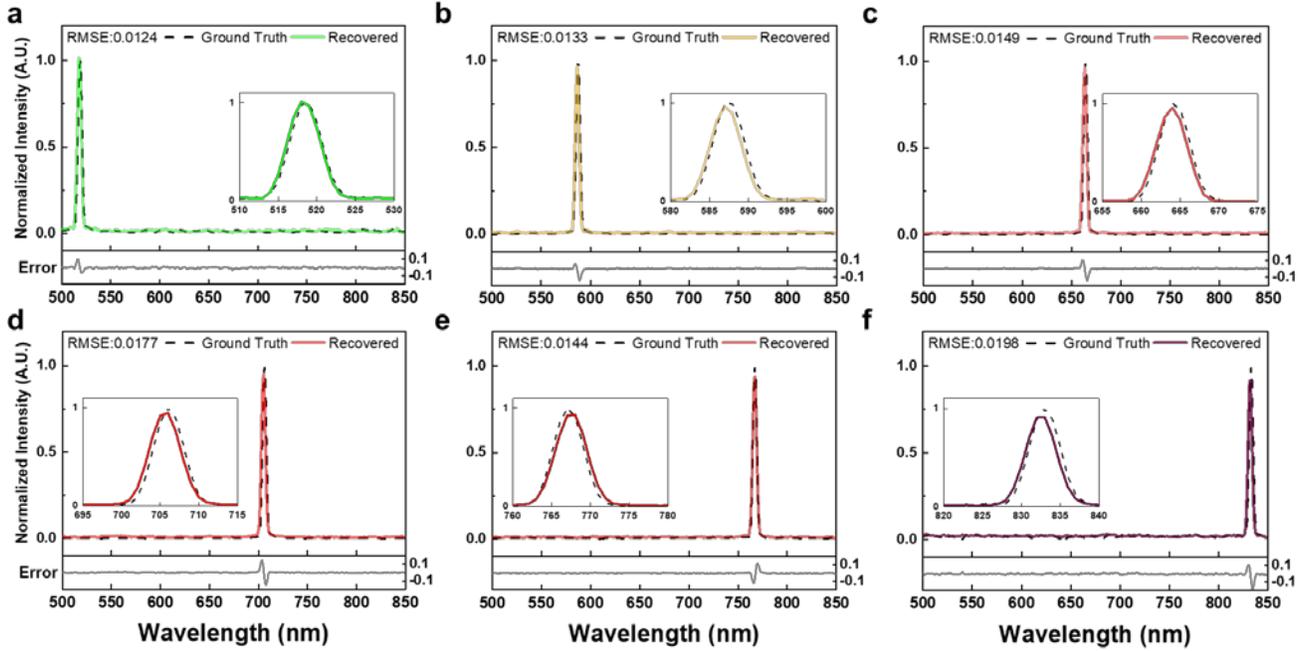

**Figure 4.** Spectral reconstructions of narrow spectra in the test set. Solid colored lines represent recovered spectra. Dashed black lines represent GT spectra. (a) 520 nm. (b) 588 nm. (c) 655 nm. (d) 707 nm. (e) 768 nm. (f) 834 nm. The RMSE between the reconstructed and GT spectrum is written in the upper left corner of each graph. Solid light gray lines at the bottom of each graph represent the error.

To train the DL architecture, we collected a dataset comprising 3,223 pairs of GT spectra and corresponding monochrome images. The dataset included 350 narrow spectra with an FWHM of 4 nm and 2,873 broad spectra. To collect narrow spectra, we used a halogen lamp (KLS-150H-LS-150D, Kwangwoo) and a monochromator (MMAC-200, Mi Optics). A beam from the halogen lamp was fed into the monochromator, generating a narrow spectrum. By changing the peak locations of narrow spectra from 500 to 849 nm with 1-nm spacing, we measured 350 spectra. To collect broad spectra, we generated various shapes of spectra using color filters (Color filter booklet, Edmund optics) as shown in Fig. 1(a). A beam from the halogen lamp was modulated by color filters, generating a broad spectrum. By changing combinations of color filters, we measured 2,873 broad spectra of various waveforms.

The CMOS camera and halogen lamp were calibrated to extract intensities from monochrome images in the range of 0–255 (Fig. 2(c)), and the CMOS camera's auto contrast function was turned off. The spectra were measured with a fixed integration time. We randomly divided the dataset into training, validation, and test sets containing 2576, 324, and 323 pairs. Before training the DL architecture, we performed data preprocessing. The measured intensities from a monochrome image were divided by the maximum value of intensities, and the corresponding GT spectrum was min-max normalized. Therefore, we trained the DL architecture to reconstruct the unknown spectra in the normalized intensity form.

We used the Adam optimizer[37] to train the DL architecture. Using the validation set, we monitored the performance of the DL architecture for every epoch during the training process. As such, we could select the number of epochs before the overfitting. The training process was completed within ~ 1.4 h, and reconstruction results on the test set came out within ~ 2 s. The DL architecture was built on the PyTorch framework[38]. The training and testing were performed on an Intel Core i7-5820K CPU computer with an NVIDIA GeForce RTX 2060 graphics processing unit.

**Reconstructions of test spectra.** Figure 3 shows the reconstruction results of the proposed computational spectrometer. The RMSE distribution of 323 test spectra is as shown in the histogram in Fig. 3(a). Blue and orange boxes represent the RMSE distribution of 33 narrow and 280 broad spectra, respectively. Three examples of the best and worst spectral reconstructions are shown in Figs. 3(b, c), respectively. Dashed black lines represent GT spectra and solid colored lines represent the reconstructed spectra. The RMSE value is written in the upper left corner of each graph. The error, defined as $\mathbf{x}_{recovered} - \mathbf{x}_{GT}$, is plotted at the bottom of each graph.

The average RMSE of all test spectra was 0.0288. The average RMSEs of narrow and broad spectra were 0.0158 and 0.0303, respectively, which indicate better results in the reconstruction of narrow spectra. As shown in Fig. 3(b), the reconstructed spectra of the best examples were almost the same as ground truth spectra. The reconstructed spectra followed well abrupt changes of narrow peaks, and peak positions match well. Moreover, the reconstructed spectra of the worst examples could not follow the waveform changes of the GT spectra well. Excluding the best and worst cases, the DL architecture recovered the test spectra well as shown in the RMSE distribution in Fig. 3(a).

Figure 4 shows spectral reconstructions of narrow spectra with an FWHM of 4 nm in the test dataset. We evenly present the reconstruction results from the test set according to peak locations. The peak locations of GT spectra in Figs. 4(a–f) are 520, 588, 655, 707, 768, and 834 nm, respectively. Solid colored lines represent the reconstructed spectra, and dashed black lines represent GT spectra. The RMSE values of the reconstructions are 0.0124, 0.0133, 0.0149, 0.0177, 0.0144, and 0.0198, respectively. The reconstructed spectrum shows spectral features of narrow peaks with drastic changes in intensities near the peak location and no intensities except at the peak location. The reconstructed spectra caught up with the steep increment of narrow peaks from the enlarged inset graphs. The peak location differences between the reconstructed and GT spectra are within 1 nm, and the FWHMs of the reconstructed spectra are within 5 nm. The proposed DL architecture reconstructed narrow spectra well regardless of the peak location.

Figure 5 shows spectral reconstructions of broad spectra in the test set. According to the interval of the histogram (Fig. 3(a)), we present the reconstruction results of broad spectra. Solid orange lines represent reconstructed spectra, and dashed black lines represent GT spectra. The RMSE values of the reconstructions in Figs. 4(a–f) are 0.0164, 0.0231, 0.0324, 0.0412, 0.0469, and 0.0523, respectively. The reconstructed spectra match well with spectral features of the GT spectra. For example, a broad background band with multiple peaks is well-expressed in Figs. 5(a-c), and spectral valleys are well-represent in Figs. 5(d, e). The reconstruction of the flat-top shape in Fig. 5(f) matches well with the GT spectrum. In addition, from the error, the differences between the reconstructed and GT spectra are within 0.1. Overall, the proposed DL architecture represents various spectra features of broad spectra.

From Figs. 3–5, we demonstrate the spectral reconstruction performances of the proposed DL architecture. The DL architecture can recover narrow and broad spectra in fine detail. In particular, it could be overfitted to broad spectra due to the different proportions of narrow and broad spectra, but narrow spectra could be well represented through the deep depth of layers and numerous learnable parameters of the DL architecture.

Unlike numerical optimization methods that require spectral sensitivities, such as transmission functions and sparsifying basis, the proposed DL architecture does not require prior information to recover unknown spectra. The DL architecture requires the dataset for training, but the DL architecture gives the reconstruction result end-to-end after the training. This is a major advantage over numerical optimization methods that require human intervention for precise parameter tuning to spectral reconstruction.

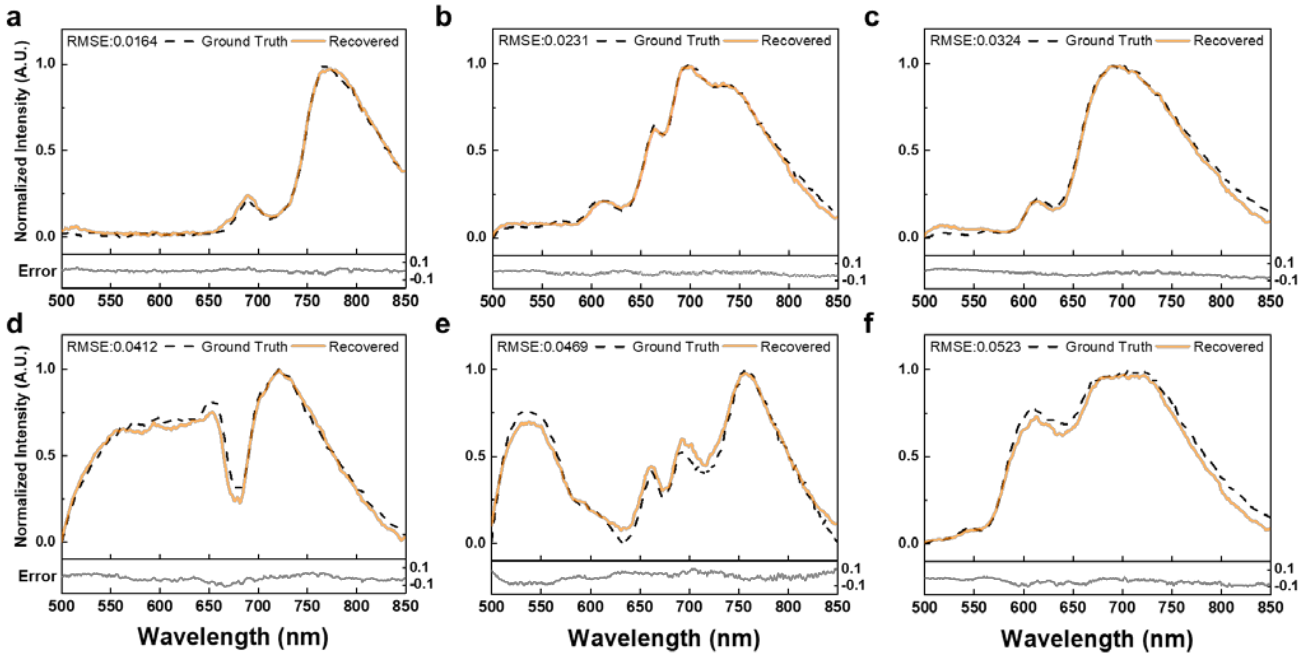

**Figure 5.** Spectral reconstructions of continuous spectra in the test set. Solid orange lines represent the reconstructed spectra. Dashed black lines represent the GT spectra. (a–f) An example of the spectrum at the second, third, fourth, fifth, and sixth intervals of the RMSE histogram, respectively. Solid light gray lines at the bottom of each graph represent the error.

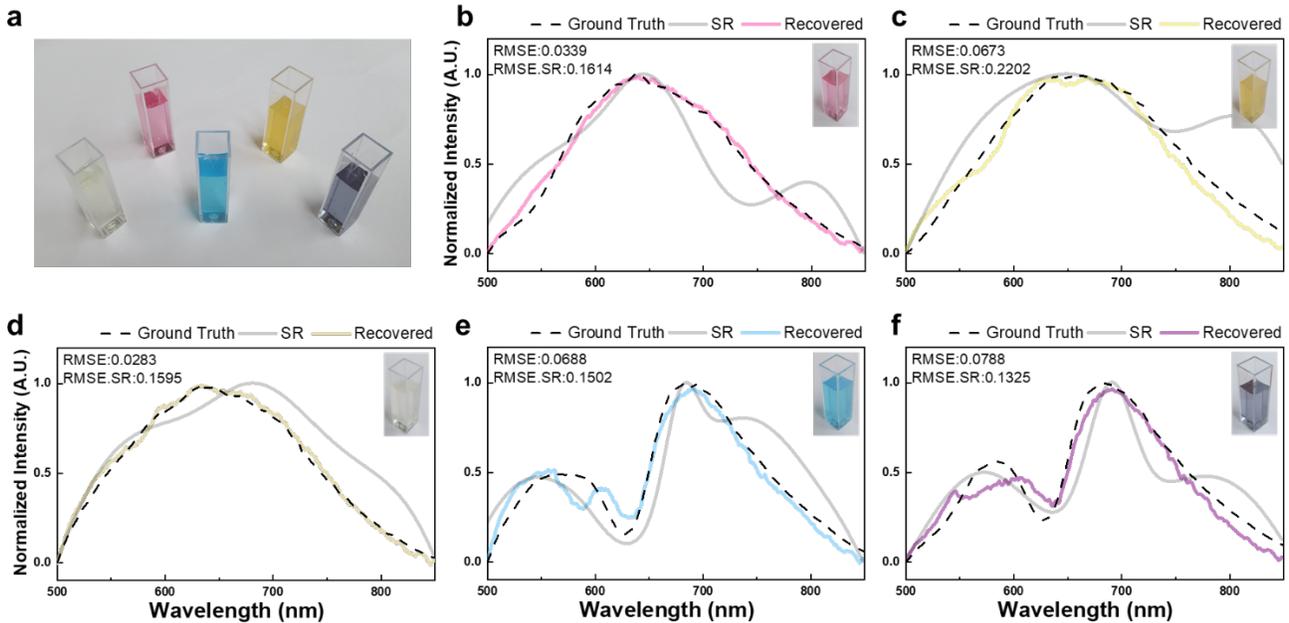

**Figure 6.** Reconstructions of the transmission spectra of five drinks. (a) Photograph of drink samples of five colors: pink, yellow, light yellow, blue and purple. (b–f) Reconstructed transmission spectra of drinks. Dashed black lines represent GT spectra, and solid lines except for light gray represent the reconstructed transmission spectra using the DL architecture. Solid light gray lines represent the reconstructed transmission spectra using the numerical optimization method, sparse recovery (SR).

**Reconstructions of transmission spectra of drinks.** We further explored the spectral resolving ability of the proposed computational spectrometer with commercial drinks. Reconstructions of transmission spectra of five drinks were performed using the trained DL architecture. The samples of five drinks were prepared using disposable polystyrene cuvettes capacity of 4 ml (Fig. 6(a)). The monochrome images and transmission spectra of drinks were measured using the experimental setup depicted in Fig 1(a) by replacing color filters with drink samples. From the monochrome image, we extracted 36 intensities and fed them into the trained DL architecture, obtaining the reconstruction result. The reconstructed transmission spectra of five drinks are illustrated in Figs. 5(b–f). Dashed black lines represent GT spectra, and solid lines represent the reconstructed spectra except for light gray. Solid light gray lines represent the reconstructed transmission spectra using the numerical optimization method, sparse recovery.

We only used the trained DL architecture without human intervention to reconstruct the transmission spectra. On the other hand, we required prior information on spectral sensitivities, the best fit sparsifying basis, and numerous interventions to find the best parameters for sparse recovery. The RMSE for each drink is written in the upper left corner of the graph. The average RMSE of the reconstructed transmission spectra using the DL architecture is 0.0554. The average RMSE using sparse recovery is 0.1648. As shown in Fig. 6, the reconstruction results of the DL architecture match well with the GT spectra. However, the reconstructed spectra of sparse recovery significantly differed from the GT spectra. The difference in reconstruction performances between the DL architecture and sparse recovery appears to be due to background noise. Because the DL architecture is trained using data with background noise, it shows stable reconstruction performances over the noise. However, sparse recovery is sensitive to noise and works well for precisely measured intensities.

Although there is still room for improvement, we demonstrate that the proposed computational spectrometer is applicable to the reconstruction of transmission spectra of drinks. Because complex optical components or long light paths are not required, the optical system (Fig. 1(a)) can be made compact. The MTF filter array can be mass-produced at a low-cost through wafer-level stencil lithography processing so that it can be fabricated as a cost effective and compact sensor for mobile applications, such as on-site detection and simple diagnostic tests.

## Discussion

The footprint of the proposed spectrometer is compact by attaching the MTF filter array to the CMOS camera. This is useful for mobile applications as it measures all filtered intensities in a single exposure. The MTF filter array can be mass-produced at a low cost by wafer-level stencil lithography processing. Unlike the conventional bandpass filters that transmit a specific light wavelength range, the MTF filters transmit light in a wide wavelength range, so the filters have a good light efficiency. In addition, because the MTF filters can offer spatial information, the MTF filter array can be developed into a snapshot hyperspectral imaging system.

We proposed a DL architecture comprising a dense layer and a U-Net backbone with residual connections to reconstruct unknown spectra. The dense layer extended the input signal size fed into the U-Net backbone, thereby deepening the U-Net backbone's depth. Therefore, the feature extraction and reconstruction could be performed in multiple steps. In the U-Net backbone, we use residual connections to prevent the gradient vanishing problem which can prevent the updating of learnable parameters. We use convolution layers with stride 2 rather than max-pooling operations for the downsampling. This replacement could improve the performance of DL architectures[39]. Finally, we summed the extended signal and output of the U-Net backbone to be a reconstructed spectrum, making the U-Net backbone learn the *residue* between the extended signal and output of the U-Net backbone.

Although previous works use numerical optimization methods to reconstruct the spectra, the demonstrations of reconstruction have been confined to monochromatic lights and LEDs. Because a target spectrum is reconstructed with a combination of sparsifying basis columns, the spectral features that can be reconstructed are limited. Therefore, the numerical optimization methods highly rely on the prior information on spectral sensitivities and the fixed sparsifying basis for precise reconstructions. However, the proposed DL architecture comprises numerous learnable parameters that can reconstruct various spectral features. In addition, the proposed DL architecture reconstructed 323 test spectra within ~2 s, which is impossible using numerical optimization methods.

## Conclusion

In this study, we developed a DL-based single-shot computational spectrometer using MTFs. The spectrometer was built using a 6 × 6 MTF filter array and a CMOS camera. The spectrometer has a compact size and fast measuring time by the single-shot structure. Using stencil lithography techniques, the MTF filter array can be mass-produced at a low-cost.

We demonstrated reconstructions of narrow and broad spectra with high resolution in a wide wavelength range (500–850 nm with 1-nm spacing) using the DL architecture for the MTF filter array-based spectrometer. The DL architecture was trained using 2,576 pairs of data in an end-to-end manner, and it took ~ 1.4 h to train the DL architecture. The trained DL architecture reconstructed 323 test spectra with an average RMSE of 0.0288 in ~ 2 s. In addition, we further explored the reconstructions of transmission spectra of commercial drinks. The DL architecture reconstructed transmission spectra consistent with GT spectra.

The proposed computational spectrometer has a compact size and fast measuring time as hardware features, high-resolution reconstruction, fast reconstruction time, and a wide working wavelength range as software features. Because of these features, the proposed spectrometer could be used in detection applications, such as drink inspection, counterfeit document detection, and self-diagnosis.

## Methods

**Fabrication of the MTF filter array.** We used TiO2 and SiO2 to manufacture the MTF filter array. TiO2 and SiO2 films were deposited onto a borosilicate glass wafer. Shadow masks were used to separate where the material should be deposited. TiO2 of desired thickness was deposited at a target location via direct current magnetron sputtering. For TiO2 deposition, a Ti target was sputtered in an Ar-and-O2-mixed gas flow of 188-sccm Ar and 12-sccm O2 with direct current power at 700 W. The TiO2 deposition was performed only on the desired area using the shadow mask. Using the other shadow mask with different patterns, we deposited SiO2 of the desired thickness. Radiofrequency magnetron sputtering was used for SiO2 deposition. A Si target was sputtered in a mixture of Ar and O2. A mixed gas flow of 185-sccm Ar and 15-sccm O2 with radiofrequency power at 300 W was used. The deposition was repeated 17 times by changing the shadow mask and materials.

**Code availability.** Custom codes for data preprocessing and the DL architecture are available from the corresponding author upon reasonable request.

## Data availability

Raw datasets and preprocessed datasets of monochrome images and ground truth spectra are available from the corresponding author upon reasonable request.

## Acknowledgements

This work was supported by a National Research Foundation of Korea (NRF) grant funded by the Korean government (MSIP) (NRF-2021R1A2B5B03002118).

## Author contributions statement

C.K. and H.-N.L. conceptualized the idea. C.K. designed and characterized the MTF filter array. C.K. constructed and conducted the optical experiment setup. C.K. and D.P. developed the DL architecture for spectral reconstruction. C.K. visualized the data under supervised H.-N.L. All authors contributed to technical discussions and writing the paper.

## Conflict of interest

The authors declare no competing interests.